\documentstyle[12pt,citews]{article}

\newcommand{\bcn}{\begin{center}}
\newcommand{\beq}{\begin{equation}}
\newcommand{\beqn}{\begin{eqnarray}}
\newcommand{\ecn}{\end{center}}
\newcommand{\eeq}{\end{equation}}
\newcommand{\eeqn}{\end{eqnarray}}

\newcommand{\sect}[2]{\vspace*{6mm}\hspace*{-\parindent}{\bf #1.}~{\bf
#2}\vspace*{4mm}}
\newcommand{\subsect}[2]{\vspace*{5mm}\hspace*{-\parindent}{\bf #1}~{\it
#2}\vspace*{3mm}}
 \def\lsim{\mathrel{\rlap{\lower4pt\hbox{\hskip1pt$\sim$}}
    \raise1pt\hbox{$<$}}}
\def\slash#1{\setbox0=\hbox{$#1$}#1\hskip-\wd0\hbox to\wd0{\hss\sl/\/\hss}}
\begin{document}
\baselineskip=18pt
\rightline{OKHEP-93-15}
\bcn
{\bf Dynamical Generation of Fermion Mass Hierarchy}
\vspace*{1cm}

{\bf Pankaj Jain }\\

{\it Department of Physics and Astronomy\\
     University of Oklahoma\\
     Norman, OK 73019}\\
\vspace*{0.5cm}

and\\

\vspace*{0.5cm}
{\bf Douglas W. McKay}\\

{\it Department of Physics and Astronomy\\
     University of Kansas\\
     Lawrence, KS 66045-2151}\vspace*{1cm}

{\bf ABSTRACT}
\ecn
\hspace*{3pc}
\parbox{30pc}{\small }We describe a simple mechanism to break  electroweak
symmetry dynamically as well as to generate a large fermion mass
hierarchy. The mechanism is displayed within the framework of
 a simple left-right symmetric model.
 The model has exact horizontal symmetry
which is broken dynamically giving rise to the mixing angles as well
as to the fermion masses. The light fermions acquire masses by exchange of the
charged gauge bosons, which is possible in the case
of a left-right symmetric gauge group. We display a simple
mechanism to break the left-right symmetry. The stability of the
horizontal symmetry breaking solution is also discussed. We argue that
if the gauge bosons are massive then for a certain range of the
gauge couplings the horizontal symmetry breaking solution may be
locally stable. We discuss some phenomenological difficulties, in particular
the question of unobserved Goldstone bosons, that arise within the present
model.
\vfill
\eject

\sect{1}{Introduction}

Construction of realistic models that implement the idea that the origin of the
fermion masses is dynamical has been pursued for many years
\citenum{{DSM7080},{summ},{summ1},{summ2},{summ3},{summ4},{summ5},{summ6}}. In
the standard model the fermions acquire masses at the tree approximation level
due to
their coupling to the Higgs particle. This has the undesirable feature
that it introduces many undetermined parameters and no explanation for
the large hierarchy in the values of the masses. Furthermore if a
similar mechanism is used in a grand unified theory it leads to acute
fine tuning difficulty \citenum{fine}. A dynamical generation of the fermion
masses
can potentially avoid all of these difficulties. Despite the
attractiveness  and the great deal of work that has been devoted to this idea
it has not so far been possible to
construct a satisfactory dynamical model which can replace the
standard model.  The main difficulty lies in generating the
experimentally observed large fermion mass hierarchy.

An attractive possibility that has recently been
suggested ~\citenum{{topcond1},{topcond2},{topcond3},{topcond4}} is that
the top quark itself may be responsible for the dynamical breakdown
of the $SU(2)_L$ $\times$ $U(1)$. The idea is that the top quark may have
some new interactions which give it a large enough condensate to
 yield the expected value for the W and Z masses.
In Ref. ~\citenum{topcond2}, the authors introduced a Nambu
Jona Lasinio interaction which couples strongly only to the third generation
of quarks. It is clearly of interest to construct a gauge theory
which incorporates the idea of top condensation. There has been
considerable work in this direction ~\citenum{topcond4}. While it is easy to
construct
a gauge interaction which can generate a large condensate for the
top quark,  generation of
 masses for the lighter fermions has proved difficult.

In the present paper we argue that the large fermion mass hierarchy
may also be generated dynamically. In other words it is also possible
to generate dynamically a very large mass for some of the fermion
generations while keeping the remaining generations massless.
We display this mechanism in section 2 by assuming that the lagrangian has an
exact or approximate horizontal symmetry. We show that,
besides the horizontal symmetric solution, the
Schwinger-Dyson (SD) equation for the fermion masses also
admits solutions which break this symmetry. In the limit
that the lagrangian has an exact horizontal symmetry, we
find a set of solutions for which only one fermion generation picks up a
nonzero mass and all other fermions remain massless. Small horizontal symmetry
breaking terms can then in principle
generate masses for the remaining fermions.
The possible Goldstone bosons which might appear because of
dynamical horizontal symmetry breaking will pick up masses once
the symmetry breaking terms are introduced into the Lagrangian.
However, as discussed in section 2, simple
examples of symmetry breaking terms do not lead to any mass
generation for the lighter fermions. We do not know if this feature
will persist for all possible symmetry breaking terms since we have
only investigated some simple examples. Rather than pursue this question here,
we investigate
 an attractive, alternative mechanism to generate masses for the
light fermions which avoids the above difficulties.

This alternative, discussed in section 3, follows naturally within the
framework of a left-right symmetric model and does
not require the introduction of any explicit symmetry breaking terms. The
approach is based on the observation that within the framework of
such models the mixing angles can also be generated
dynamically. This can be accomplished since the matrix structure of the
dynamically generated up quark mass matrix does not have to be the
same as the down quark mass matrix. Indeed for the case of a 2
generation model there exists an $SU(2)$ degeneracy for both the
up quark sector and the down quark sector if the horizontal symmetry is
spontaneously broken. This degeneracy allows us to choose different
matrix structure for the up and down mass matrices and therefore allows
us to generate mixing angles dynamically. Once the mixing angles are
generated, then within the framework of a left-right symmetric model the
lighter fermions naturally acquire masses by higher loop diagrams.

 Several possible
objections to such a mass generation mechanism are also addressed.
An important issue is the question of stability of the horizontal symmetry
breaking solutions. In order to have chiral symmetry breaking we require
strong coupling, and since we have horizontal symmetry at the lagrangian
level all fermions are coupled to this strong interaction. Therefore it
might be argued that the solution with some massless fermions will be unstable.
However the point is that the gauge boson that generates this strong
interaction is assumed to acquire a very large mass dynamically. Therefore
for light fermions this interaction acts like a very weak interaction although
the coupling strength is large. This suggests that the solution with some
massless fermions will certainly be locally stable. In other words it may be a
local minima of the effective potential and might have a higher energy than
the global minima. The proper definition of the effective potential within the
present framework as well as its limitations are discussed in section 2. Our
physical picture is that the effective potential has several local minima and
during the early universe as the chiral symmetry breaking phase transition
takes
place the universe can choose any one of these minima. This is allowed as long
as the transition time for the universe to go from this minima to another,
which might be lower in energy, is longer than the age of the universe. In the
present
paper, we do not calculate the transition time since within the dynamical
framework such a calculation is very nontrivial. Therefore we only demonstrate
that the solution under consideration
is locally stable and leave a proper discussion of decay rate of this solution
for further research.

Finally we also discuss the important problem of the possible existence of
Goldstone bosons if the horizontal symmetry is broken dynamically. This problem
is, however, not resolved in the present paper. We argue that although some of
these bosons will acquire mass because of the anomalous nature of the
corresponding broken
symmetry, a number of these bosons will remain massless. In particular for the
case of two generations because of the dynamical breakdown of $SU(2)_L\times
SU(2)_R$, there will appear six Goldstone bosons. It is clearly necessary to
avoid these zero-mass particles in order that the
present scheme may be phenomenologically viable. We hope that this can be
accomplished by a suitable extension of the present model, which might involve
gauging the horizontal symmetry with the gauge coupling being very weak so as
not to play any role in the low energy physics. We are currently investigating
different possibilities which might avoid the Goldstone bosons.

\sect{2}{Basic idea}

To display the basic mechanism
we construct a simple model with two fermions.  The
fermions have a nonabelian SU(3) interaction -namely, Quantum Chromodynamics
(QCD). We ignore the electro-
weak interactions for the present discussion. The application of
our ideas to the standard model will be discussed
in sect. 3. We do not
introduce any fundamental scalar fields. This model has a global $SU(2)_L\times
SU(2)_R$ horizontal symmetry besides the color gauge symmetry and the global
$U(1)_V$ symmetry. The lagrangian for the
theory can be written as,
\beq
{\cal L} = -{1\over 2} \mbox{tr}\ G^{\mu\nu}G_{\mu\nu} + \sum_{a=1,2}
i\bar\psi_a\slash \partial\psi_a +
g\sum_{a=1,2} \bar\psi_a \slash A\psi_a
\eeq
where $$G_{\mu\nu}=\partial_\mu A_\nu-\partial_\nu A_\mu - ig[A_\mu,A_\nu]$$
$$A_\mu = \sum_{i=1}^8 A_\mu^i\lambda^i/2$$ and $\lambda^i$ are the generators
of $SU(3)$. The two fermions, corresponding to $a=1,2$, are assumed to have all
charges identical and are analogous to, for example, the strange and down
particles. The horizontal $SU_L(2)\times SU_R(2)$
symmetry corresponds to the unitary transformations,
\beq
\left(\matrix {\psi_1'\cr \psi_2'\cr}\right)_L = U \left(\matrix {\psi_1\cr
\psi_2\cr}\right)_L\ \ \ ,\ \ \ \left(\matrix {\psi_1'\cr \psi_2'\cr}\right)_R
= U \left(\matrix {\psi_1\cr \psi_2\cr}\right)_R
\eeq
We assume that chiral symmetry is dynamically broken in this
model because of the QCD-like interaction. The propagator for this fermion
doublet can be written as
\beq
S(k) =	i \bigg[A(k^2)\slash k - M(k^2)\bigg]^{-1}
\eeq
where $A(k^2)$ and $M(k^2)$ are two by two matrices, assumed to be symmetric
for simplicity,
 which can be
determined by solving the SD equation, which can be written in the ladder
approximation as,
\beq
S^{-1}(q)\ =\ S_0^{-1}(q) + \int {d^4k\over (2\pi)^4}
g^2\gamma_\mu S(k)\gamma_\nu\ G_{\mu\nu}(k-q)
\eeq
\noindent where $S_0(q)$ is the bare quark propagator, assumed to have no mass
term, and $G_{\mu\nu}$ is the gluon propagator given by
\beq
G_{\mu\nu}(k) = -i\bigg(g_{\mu\nu}-{k_\mu k_\nu\over k^2}\bigg)\ G(k^2)
\eeq
We use the Landau gauge for our calculation and
for simplicity will assume
that $A(k^2)$ is equal to the identity matrix. This should not have
any qualitative effect on our results since model calculations in the single
fermion case show that $A$ differs from 1 by no more than a factor 2
\citenum{afunc}. Our objective is to obtain
a nonperturbative chiral symmetry breaking solution for $M$ by
solving this equation. The equation satisfied by $M(q^2)$, assumed
to be symmetric, is given in Euclidean space by,

\beq
M(q^2) = -3\int{d^4k\over (2\pi)^4}\ g^2G(k-q)\ M(k^2)\ \bigg[k^2
+M^2(k^2)\bigg]^{-1}
\eeq
This leads to the following equations for the
 matrix elements  $m_{11}(q^2)$, $m_{22}(q^2)$ and
$m_{12}(q^2)$ of $M$,
\beqn
m_{11} &=& -3\int {d^4k\over (2\pi)^4} g^2G[(k-q)^2]{m_{11}k^2 + m_{22}\ D_{-}
\over k^4+k^2D_{+}+D_{-}^2} \nonumber\\
m_{22} &=& -3\int {d^4k\over (2\pi)^4} g^2G[(k-q)^2]{m_{22}k^2 + m_{11}\ D_{-}
\over k^4+k^2D_{+}+D_{-}^2}	\nonumber\\
m_{12} &=& -3\int {d^4k\over (2\pi)^4} g^2G[(k-q)^2]{m_{12}k^2 - m_{12}\ D_{-}
\over k^4+k^2D_{+}+D_{-}^2}
\eeqn
where
\beq
	D_{-} = m_{11}m_{22} - m_{12}^2\nonumber
\eeq
and
\beq
	D_{+} = m_{11}^2+m_{22}^2 + 2m_{12}^2\ \ .\nonumber
\eeq

It turns out that there are several different chiral symmetry-breaking
solutions of this equation. The usual choice is to assume that the
mass matrix is of the form,

\beq
M(q^2) = m(q^2)\ I
\eeq
where $I$ is the identity matrix.
However, another possibility is a solution for which one of the
eigenvalues is zero. In other words there exists a solution to Eq. 6 of the
type,
\beq
M(q^2) = m(q^2) \left(\matrix{1&0\cr 0&0\cr}\right)
\eeq
To our knowledge, this type of solutions have not been discussed in the
literature within a fully dynamical mass generation framework.
This solution can of course be interesting only if it is stable, at least under
local fluctuations. By local fluctuations we mean small oscillations around the
minimum of the effective potential. An appropriate framework for the
construction of effective potential within the present context is discussed
below. If the solution is not a global minima then we have to require that its
decay time is longer than the age of the universe. We postpone the discussion
of stability until section 3.
By substituting the ansatz given in Eq. 11 into Eq. 6 one finds,
\beq
m(q^2)\ =\ -3 \int {d^4k\over (2\pi)^4}\ g^2\ G(k-q){m(k^2)\over
k^2+m^2(k^2)}
\eeq
which is expected \citenum{sdeqn} to have a chiral symmetry breaking solution.
Simple models for $G(k^2)$ which lead to such solutions in the ladder
approximation are discussed in Ref. \citenum {modelG}. The ansatz in
Eq. 11 therefore naturally generates a large hierarchy in masses between the
two fermions. Since this solution spontaneously breaks the horizontal symmetry
we can generate an entire class of solutions by making a unitary transformation
on the ansatz in Eq. 11.
\beq
M\rightarrow U^\dagger M U
\eeq
where $U$ is an $SU(2)$ matrix. Any (det$M=0$) solution \citenum{ansatz} of
this type is clearly a solution to the SD equation since the original
lagrangian is
symmetric under this
transformation. Furthermore we expect that if we have a well
defined measure of energy density corresponding to these solutions then these
solutions
will be degenerate. We can see this by looking at
the Cornwall, Jackiw and Tomboulis (CJT) effective potential \citenum{CJT}
\beq
V(S) = -i\int {d^4p\over (2\pi)^4} Tr\bigg\lbrace\ln\left[(S^0)^{-1}S\right] -
(S^0)^{-1}S
+1\bigg\rbrace -{i\over 2}\int{d^4p\over (2\pi)^4} Tr\Sigma(p)S(p)\ \ ,
\eeq
where
$$\Sigma(q) = (1-A)\slash q+M = M = \int{d^4k\over (2\pi)^4} g^2\gamma_\mu
S(k)\gamma_\nu G_{\mu\nu}(k-q) $$
when the ladder approximation and $A=1$ assumptions are made. $V(S)$ is written
in a form such that the SD equation ${\delta V\over \delta S}=0$ leads
transparently to Eq. 4 \citenum{stam}. The degeneracy of solutions related to
(11) by unitary transformations (13) can be seen immediately from Eq. (14).
Since $\Sigma=M$ and $S^{-1}=\slash q+M$ in our approximation, it is clear that
the transformation (13) does not change the effective potential.
This suggests that the eigenenergies for these
modes of fluctuation will be zero, signalling the
appearance of three Goldstone bosons corresponding to spontaneous
breakdown of $SU(2)$ horizontal symmetry.
Although the above discussion is reasonable,
 there remain difficulties as to the proper
interpretation of the effective potential \citenum{banks} and therefore it is
not completely
clear as to what are the correct criteria for stability. For the present paper
we require that in order to be stable the solution to the gap equation should
not lead to any tachyonic solutions to the fermion-antifermion Bethe-Salpeter
bound state equation.
This is reasonable since the Bethe-Salpeter equation is obtained by taking
the second variation of the effective potential with respect to the fermion
propagator.
 This criteria will be used in section 3, where we present a suitable
generalization of the standard model to display dynamical fermion and gauge
boson mass generation, to suggest that in some models dynamical horizontal
symmetry breaking solutions are locally stable apart from
the zero modes discussed above.

The above model can be generalized straightforwardly to any number of fermions.
In all these cases if the
original lagrangian has perfect SU(N) horizontal symmetry then there exists a
solution for which
only one fermion acquires mass and all other fermions remain
massless.

We have so far discussed a toy model for which the lagrangian
 has a perfect horizontal
symmetry. We have found that there exists
a solution to the gap equation which breaks this
symmetry such that only one of the fermions acquires mass and
all other fermions remain massless. We next discuss the very
important issue of generating masses for these light fermions.
This can in principle
be done in the present framework by introducing
small horizontal symmetry breaking terms in the lagrangian. We
again specialize to the simple case of two fermions. The horizontal
symmetry can be broken, for example, by introducing a
new interaction which couples only to the second quark.
The strength of this coupling is assumed to be much weaker
in comparison to the dominant chiral symmetry breaking interaction. The
solution to the gap equation in the
presence of explicit horizontal symmetry breaking, however, is somewhat subtle.
The reason
for this can be explained as follows. As discussed above, in the
absence of horizontal symmetry breaking, there exists a continuum of solutions
for $M$ which lead to the same mass spectrum. These solutions
are related to one another by a unitary transformation of the
mass matrix. Since the lagrangian is symmetric under such transformations, the
effective potential for these different
solutions turns out to be equal which means that these solutions
can be deformed into one another without spending any `energy'. Therefore
in the absence of the symmetry-breaking term any one of these solutions
is as good as any other. However, in the presence of the symmetry-breaking
terms, the degeneracy between these solutions will be removed and only one
of these solutions will be selected, in analogy with the partial conservation
of axial current analysis. This solution should have the lowest
energy in the presence of the symmetry-breaking terms. Therefore before
expanding the solution in powers of the symmetry-breaking parameter it is
necessary to determine the suitable unperturbed solution in the presence of the
symmetry breaking terms.
By following this procedure the SD equation can be solved for some simple
examples.

One such example is a symmetric NJL model with coupling strength large enough
to dynamically break chiral symmetry. The explicit horizontal symmetry-breaking
interaction can also be chosen to be another NJL model which couples only to
the first quark and with a much weaker interaction strength. The lagrangian for
this model can be written as,
\beq
{\cal L} = \bar\psi i\slash\partial\psi + {g^2\over M^2}\left[(\bar\psi\psi)^2
+(\bar\psi i\gamma_5\psi)^2\right] + {g'^2\over M^2}\bigg\lbrace\left[\bar\psi
\left(\matrix{1&0\cr 0&0\cr}\right)\psi\right]^2
+\left[\bar\psi i\gamma_5\left(\matrix{1&0\cr
0&0\cr}\right)\psi\right]^2\bigg\rbrace
\eeq
where
$$\psi = \left(\matrix{\psi_1\cr\psi_2}\right)$$
and $g'<<g$.
 By explicit calculation we find that in this model, for the case of dynamical
horizontal symmetry breaking, one of the fermions remains massless even though
explicit horizontal symmetry breaking have been introduced. This happens
because in the presence of such a symmetry-breaking term the unperturbed
solution has to be chosen such that only one of the entries, either $m_{11}$ or
$m_{22}$ is
nonzero in the 2$\times$2 mass matrix. It may be possible
to choose some more general symmetry breaking terms which do lead to mass
generation for the light fermions. However, as discussed in the introduction,
for the present paper we introduce an
alternative, more attractive, approach which does not require the introduction
of any ad-hoc symmetry breaking terms. The approach is based on a
left-right symmetric model and is discussed in next section.

\sect{3a}{Explicit model}

We construct an extension of the standard model which incorporates
the dynamical mechanism for fermion mass hierarchy. In the present paper we
 illustrate that such a mechanism is possible without studying
detailed phenomenology. Construction of a realistic model and its
phenomenological consequenses will be left for further research.

In order to construct a dynamical symmetry breaking model we need
some interaction strong enough to admit chiral-symmetry-breaking
solutions. We assume that this is not possible within pure $SU(2)_L\times
U(1)$.
A detailed study of the possibility of dynamical breakdown of chiral
symmetry within $SU(2)_L\times U(1)$ has been done by Mahanthappa and Randa,
Ref. \citenum{summ6}.
 We extend the standard model to include
an $SU(2)_R$ interaction and assume that the $U(1)$ coupling strength is
large enough to break chiral symmetry. The dynamical breaking gauge
group for the present study then becomes $SU(3)_{color}\times SU(2)_L\times
SU(2)_R\times U(1)$. As discussed later, besides dynamically breaking the
electroweak gauge symmetry,
this model also leads naturally to mass generation for light fermions
without the introduction of any horizontal symmetry breaking terms.
In order to be
realistic we require that six gauge bosons become massive with
three of the gauge bosons having masses considerably heavier than the scale of
the electroweak symmetry breaking.
This can be accomplished by introducing a
 new fermion doublet which couples only to the
$SU(2)_R$ gauge bosons. This fermion doublet is assumed to have
some new interaction which leads to the formation of a condensate
much larger than the scale of electroweak symmetry breaking. As
discussed later such a model accomodates a realistic gauge boson spectrum.

The lagrangian for this theory can be written as,
\beqn
{\cal L} &=& {\cal L}_{QCD} + {\cal L}(A_L^\mu) + {\cal L}(A_R^\mu) + {\cal
L}(B^\mu)
+ {\cal L}(\psi) + {\cal L}_I + {\cal L}(\chi)\nonumber\\
{\cal L}_I &=& g_L (\bar\psi_1\ \bar\psi_2)_L\ {\bf \tau\cdot\slash A_L}
\left(\matrix {\psi_1\cr\psi_2\cr}\right)_L
+g_R (\bar\psi_1\ \bar\psi_2)_R\ {\bf \tau\cdot\slash A_R}
\left(\matrix {\psi_1\cr\psi_2\cr}\right)_R\nonumber\\
&+&\tilde gY_L (\bar\psi_1\ \bar\psi_2)_L\ \slash B
\left(\matrix {\psi_1\cr\psi_2\cr}\right)_L
+\tilde gY_R (\bar\psi_1\ \bar\psi_2)_R\ \slash B
\left(\matrix {\psi_1\cr\psi_2\cr}\right)_R
\eeqn
where $\tau_i$ are Pauli matrices, $(\psi_1, \psi_2)$ is a fermion doublet,
$A_L$, $A_R$ and $B$ are $SU(2)_L\times SU(2)_R\times U(1)$ gauge bosons and
 ${\cal L}(A_L^\mu),\ {\cal L}(A_R^\mu),\ {\cal L}(B^\mu),
\ {\cal L}(\psi)$ are the kinetic energy terms for the gauge particles
and the (massless) fermions.  ${\cal L}(\chi)$ refers to the lagrangian of
the extra fermion doublet which does not couple to the left handed particles.
An explicit lagrangian for this doublet is given in Appendix 1.
$Y_L$ and $Y_R$ are equal to $-1$ for
leptons and 1/3 for quarks. The generation index has been suppressed in the
above lagrangian. We assume that there are $N$ generations, all of which
have identical couplings to the gauge particles. Therefore the lagrangian has
$SU(N)_L\times SU(N)_R$ global horizontal symmetry,
\beq
\left (\matrix{\psi_1\cr \psi_2}\right)_{L,i}\rightarrow U_{ij}\left(
\matrix{\psi_1\cr \psi_2}\right)_{L,j}\ \ , \ \  \left (\matrix{\psi_1\cr
\psi_2}\right)_{R,i}\rightarrow U_{ij}\left(
\matrix{\psi_1\cr \psi_2}\right)_{R,j}
\eeq
To get the spectrum of this model it is helpful
to first identify the vector gauge boson which would not acquire
mass after dynamical symmetry breaking. We express the fields $A_R^3$ and
$B$ in terms of $R_1$ and $R_2$ by introducing the angle $\theta_1$,
\beqn
A_R^3 &=& \cos(\theta_1)\ R_1\ - \sin(\theta_1)\ R_2\nonumber\\
B &=& \sin(\theta_1)\ R_1\ + \cos(\theta_1)\ R_2
\eeqn
The angle $\theta_1$ is chosen such that,
\beq
\cos(\theta_1) = {\tilde g\over \sqrt{g_R^2+\tilde g^2}}\ \ .
\eeq
The interaction lagrangian then becomes
\beqn
{\cal L}_I &=& g_L (\bar\psi_1\ \bar\psi_2)_L\ (\tau\cdot\slash A_L
\left(\matrix {\psi_1\cr\psi_2\cr}\right)_L +
 \nonumber\\
&+&g_R (\bar\psi_1\ \bar\psi_2)_R\ (\tau_1\slash A_R^1+\tau_2\slash A_R^2)
\left(\matrix {\psi_1\cr\psi_2\cr}\right)_R
+ \tilde g\sin\theta_1Y_L (\bar\psi_1\ \bar\psi_2)_L\ \slash R_1
\left(\matrix {\psi_1\cr\psi_2\cr}\right)_L\nonumber\\
&+& \tilde g\sin\theta_1 (\bar\psi_1\ \bar\psi_2)_R\
\left(\matrix{1+Y_R&0\cr 0&-1+Y_R\cr}
\right)\slash R_1\left(\matrix {\psi_1\cr\psi_2\cr}\right)_R\nonumber\\
&+& (\bar\psi_1\ \bar\psi_2)_R\
\left(\matrix{-g_R\sin\theta_1+Y_R\tilde g\cos\theta_1&0\cr 0&
g_R\sin\theta_1+\tilde g\cos\theta_1Y_R\cr}
\right)\slash R_2\left(\matrix {\psi_1\cr\psi_2\cr}\right)_R\nonumber\\
&+& \tilde g\cos\theta_1Y_L (\bar\psi_1\ \bar\psi_2)_L\ \slash R_2
\left(\matrix {\psi_1\cr\psi_2\cr}\right)_L
\eeqn

The identification of the above model with the standard $SU(2)_L\times U(1)$
is now clear. The couplings of the field $R_1$ are precisely the
ones chosen for the $U(1)$ field in the standard model with $\tilde g
\sin\theta$ as the $U(1)$ effective coupling and $\pm 1+Y_R$ as the
$\psi_{1R}$,
$\psi_{2R}$ hypercharges. The
photon and the $Z$ fields can be defined by rotating the fields
$R_1$ and $A_L^3$ by the Weinberg angle. The extra interactions
in the above lagrangian involve the fields $A_R^1$, $A_R^2$ and $R_2$.
These fields will be assumed to have masses much larger than the
scale of electroweak interactions and therefore do not play any
role in the low energy physics. As mentioned above this can be accomplished by
introducing another fermion doublet which transforms only under the $SU(2)_R$
group and does not couple to the $SU(2)_L$ gauge bosons. The lagrangian for
this doublet is explicitly displayed in Appendix 1. As discussed in this
appendix, we assume that this fermion doublet has a large condensate due to a
new strong
interaction. The right handed gauge bosons then pick up large
masses proportional to this condensate. This new fermion doublet also couples
to the photon and the $Z$ boson and its hypercharge
assignment can be chosen so that the photon does not
pick up any mass. This can be accomplished by giving different left
hypercharges $Y_L^U$ and $Y_L^D$ to the left handed up and down quark partners
of this heavy right handed doublet. This is allowed since this doublet does
not couple to SU(2)$_L$. The charge has to be chosen such that,
\beq
\left(\matrix {Y_L^U\cr Y_L^D\cr} \right) = \left(\matrix {1+Y_R\cr -1+Y_R\cr}
\right )
\eeq
Therefore if $Y_R$ for this doublet is taken to be 1/3, as for all other
generations,
we get $Y_L^U=4/3$ and $Y_L^D= -2/3$. With this assignment the coupling of the
photon and the $Z$ to this doublet is purely vectorial. Therefore photon will
not acquire any mass. Furthermore the correction to the $Z$ mass will be
higher order in the weak coupling since in the absence of all other
interactions
the condensate due to this new right handed doublet cannot generate mass
for the $Z$. Therefore the correction to the $\rho$ parameter will be small.

\subsect{3b.} {Dynamical generation of mixing angles}

In order to give masses to the remaining fermions and the $W $ and $Z$ we
assume that $\tilde g$ is large enough to admit dynamically broken solutions
 to
the gap  equation. We first analyze the gap equation for this case in the
ladder
approximation, which we expect to be qualitatively reliable. The dominant
contribution is given by the $R_2$ gauge boson. We may therefore drop the
contributions due to all other gauge bosons, which can be calculated
perturbatively once the dominant term has been calculated. The resulting
gap equation in this approximation is shown diagrammatically
in Fig. 1. We remind the reader
that the fields $A_L^3$ and $R_1$ in Eq. 14 will mix to yield the
photon and the Z.  The solutions to the resulting massive gauge boson
exchange  gap equation for one generation
in the ladder approximation are discussed in many papers
\citenum{maskkugo}.
We accept that such solutions can be determined
and continue to analyze the solution with several generations.

We first specialize to the case of two generations.
As discussed in section 2, in this case there are at least two possible
types
of chiral-symmetry-breaking solutions
to the SD equation. One of these classes of solutions breaks
the horizontal symmetry also. We will choose this solution
for further analysis. The important point is that there are
many possible solutions related to one another by a unitary
transformation which break horizontal symmetry and
lead to the same final eigenvalues, only one of which is
nonzero. These different solutions however have different
eigenvectors. We can pick any one of these solutions for the
up quark sector and a different solution for the down
quark sector. As we diagonalize these mass matrices we will
generate mixing angles since the up and down mass matrices will
have to be rotated by different amounts. In this sense our model can
be said to dynamically generate the mixing angles.

We finally end up with a situation in which only one of the
quark generations has acquired a mass and the mixing angle
of this generation with the remaining generations is also
nonzero. Now we include the effects of the usual electroweak
interactions. The Feynman diagram shown in figure (2) will
generate masses for the lighter generations. This is discussed
in greater detail below.

\subsect{3c.}{Stability of the horizontal symmetry breaking solution}

It is clearly important to determine whether the horizontal symmetry
breaking solution discussed above is stable. However it is not
completely clear what is the correct criteria to determine the
stability of such solutions because of the usual difficulties in
the interpretation of the effective potential in a dynamical, bilocal
 framework as the energy density \citenum{banks}.
A possible criterion is to require that there does not exist any
tachyonic
 solutions to the bound state Bethe-Salpeter (BS) equation. This
 has been
discussed in some detail in Ref. \citenum{higashi}. This
is reasonable since the BS kernel is obtained by taking the second
derivative of the effective action with respect to the fermion propagator
\citenum{mcmn} and the BS equation spectrum should represent
the eigenmodes of fluctuation around the solution to the gap equation.
A tachyonic solution therefore will represent a possible unstable
eigenmode and a breakdown of the vacuum. For example if we consider
the $f\bar f$ BS bound state equation  for QED in the ladder approximation,
as the coupling is increased the ground state mass plunges to zero
at $\alpha=\pi/4$. If the coupling is increased beyond this value then
a
tachyonic solution may appear. However at this value of the coupling
 the
perturbative vacuum is expected to break down and chiral symmetry may be
dynamically broken. It is interesting that in the case of ladder BS
equation in  QED the
bound state mass plunges to zero at roughly the same coupling as
is necessary for obtaining the dynamically broken solution to the
corresponding gap equation in the ladder approximation, for which
the
critical value of $\alpha=\pi/3$.

 We will use this criterion to determine if the horizontal
symmetry breaking solution is stable. Specifically we demand that there
does
not exist any zero mass bound state if the vacuum is chosen to be the
perturbative one.
For simplicity we pick a
particular solution for which all elements of the mass matrix (for two
generations) are zero except one. For such a solution the two generations
decouple and we will study the stability of the two generations
separately. In other words we are ignoring the quantum fluctuations which
mix the two generations. Because of horizontal symmetry of the lagrangian,
 the behavior of the `potential' under these fluctuations is already known
 and it leads to the appearance of Goldstone bosons arising because of the
spontaneous breakdown of the horizontal symmetry.
 Therefore for the present discussion we need to only concentrate on
fluctuations within a particular generation.
 For the case of the massive generation, the fermions acquire
dynamical
masses and the vacuum will be stable
apart from the appearance of Goldstone bosons.
 However for the
generation which remains approximately massless there is a possibility
that
 a tachyonic solution might exist if the coupling is large enough.
In the present case
we are considering a massive gauge boson exchange which is
responsible for chiral symmetry breaking in the heavy generation. For the
 case
of the light generation, however, unless the coupling is much greater than
unity,
 because of the heaviness of the gauge particle,
the gauge interaction will effectively act as a very weak interaction. In
particular for a large range of coupling values this gauge interaction
may not be strong enough to even form a bound state for the light fermions.
 This can be
seen by determining the minimum fermion mass needed to form a bound
state for a Yukawa potential. It is given by,
\beq
m_{\rm fermion} > {8\over\alpha}\ M_{\rm Boson}
\eeq

The above result is obtained by using the Schrodinger equation. However,
as discussed in Ref. \citenum{jain}, several different relativistic models
give results very consistent with equation (22).
Therefore if $\alpha$ is of order 1 and if the mass of the fermion
is much smaller than the mass of the boson there will not even exist
a bound state, much less a bound state with zero $M_B^2$, which can
only arise if the coupling is much larger than is required to form bound
states with weak binding. In other words, although the gauge coupling in
the present case is large, because of the large gauge boson mass,
it behaves essentially as a weak coupling
for the fermion of very small mass.
The above argument suggests that the
horizontal symmetry breaking solutions may be locally stable except for a
finite
number of 'zero modes' corresponding to the spontaneous breakdown of the
horizontal symmetry. By locally stable we
mean that the solution is a local minima of the CJT effective potential but
may not be a global minima. In order for the solution to be realistic it
is necessary to check that the decay time for this solution is longer than
the lifetime of the universe. However within the present framework we do not
so far have a reliable methodology to calculate the decay rate because of
 the usual difficulties with the interpretation of the effective potential
 in a dynamic, bilocal framework as the energy density \citenum{banks}.

\subsect {3d.}{`Trickle-down' mechanism for light fermion mass generation}

The left-right symmetric model described above can dynamically generate
 mixing
angles as well as masses for the lighter fermions. The basic point
can be explained by considering a 2 generation model. As discussed
above there exists an infinite family of horizontal symmetry breaking
solutions. All of these solutions are related to one another by
a $2\times 2$ unitary transformation. We have the freedom to pick any
 one of
these solutions for the up quark sector and a different one for the
down quark sector. The up and down quark solutions are different in terms
of their matrix structure in the flavor space. These two solutions both
lead to the fermion
mass hierarchy but require different unitary transformations
for diagonalization. Such a choice of solutions will therefore
dynamically generate mixing angles. Once the mixing angles are
generated then higher order diagrams \citenum{pagels} like the one shown
in Fig. (2) will let us generate nonzero masses for the lighter generations,
as we now demonstrate.

By the above reasoning we know that to leading order we
get a $2\times 2$ Cabibbo, Kobayashi-Maskawa (CKM) matrix of the form
$$\left(\matrix{V_{11}&V_{12}\cr -V_{12}&V_{22}}\right)\ \ .$$
The model does not predict the value of the mixing angle since
all possible values are allowed and the correct one has to be determined by
 experiment.  It is possible that if the charged particle exchange is
treated nonperturbatively, CKM matrix may be uniquely specified. However, we
have not
 so far been able to develop a simple approximation scheme to perform
 this calculation and for the present paper will assume a general form
for this matrix. For the sake of illustration the heavy
generation will be assumed to be a hypothetical fourth generation (T,B)
with mass scale of the order of the $R_2$ gauge particle. The
light generation will be taken to be the third generation.
Including the perturbative correction due to the diagram shown in Fig. (2)
we
find that the resulting up and down sector mass matrices are given by,
\beqn
M_U &=& \left(\matrix{m_{T0}+V_{11}^2I_B&-V_{11}V_{12}I_B\cr -V_{11}V_{12}I_B
&V^2_{12}I_B\cr}\right)\nonumber\\
M_D &=& \left(\matrix{m_{B0}+V_{11}^2I_T&V_{11}V_{12}I_T\cr V_{11}V_{12}I_T
&V^2_{12}I_T\cr}\right)
\eeqn
where, $m_{T0}$ and $m_{B0}$ are the fourth generation top and bottom masses
calculated with an appropriate truncation of the SD equation,  which is taken
to be the ladder approximation for the present study. The dominant contribution
is
obtained by including only the $R_2$ gauge boson. The corrections due to the
photon and the Z particles are suppressed by one power of the weak coupling and
will be ignored for the present discussion.
$I_T$ and $I_B$ represent the integrals
corresponding to figure 2 with the internal fermion propagator
being the fourth generation top and bottom respectively. These integrals
 are considerably smaller than $m_{T0}$ and $m_{B0}$ because of the weak
coupling factors.
 The mass matrices for the up and down sector particles can be
diagonalized by the usual procedure by transforming the fields by a unitary
matrix. The correction to the CKM matrix can then be calculated. The rotation
angles $\theta_U$ and $\theta_D$ which diagonalize the mass matrices displayed
in equation 23 are given by,
\beqn
\tan2\theta_U &=& {-2V_{12}V_{11}I_B\over m_{T0}+V_{11}^2
I_B-V_{12}^2I_B}\nonumber\\
\tan2\theta_D &=& {2V_{12}V_{11}I_T\over m_{B0}+V_{11}^2I_T-V_{12}^2I_T}
\eeqn
The above result follow by including only the charged particle exchange.
All other particle exchanges give negligible contributions.
Equation 24 shows that
$\theta_U$ and $\theta_D$, since they are proportional to $I_B$ and $I_T$, give
perturbatively small corrections to the CKM matrix. The diagonalizations of the
up and down quark mass matrices lead to the following mass eigenvalues,
\beqn
m_B \approx m_{B0}\ \ &,&\ \ \ m_b \approx V_{12}^2 I_T\nonumber \\
m_T \approx m_{T0}\ \ \ &,&\ \ \ m_t \approx V_{12}^2I_B
\eeqn
The above equations reveal that the
third generation top and bottom quark masses are proportional to the small,
perturbative corrections $I_B$ and
$I_T$ respectively and therefore we have generated masses for the third
generation which are much smaller than the fourth generation. Thus we have
dynamically generated a mass hierarchy between two generations.

The above result is clearly promising but in the present form either our model
or the assumed solution to the SD equation is not phenomenologically
acceptable.
 Although we have obtained a mass hierarchy between the third and the fourth
generations, the mass hierarchy within the third generation has not been
accomplished. We see this by considering $I_B$ and $I_T$ which are proportional
to $m_B$ and $m_T$. It can
be seen from the lagrangian that $m_B$ and $m_T$ cannot be too different
since the effective couplings of the up and down quark sectors are roughly the
same. This implies that the fourth generation top and bottom are roughly
degenerate. However this along with Eq. 25 implies that the third generation
top and bottom are also degenerate, in conflict with experiments. So the model
as it stands is not able to give a realistic mass spectrum. This problem has
clearly
arisen since the fourth generation is approximately degenerate and feeds its
degeneracy down to the lower generation.

A possible way out of this problem is to choose a different dynamically broken
solution. We may choose the solution in which the up quark mass matrix does not
break horizontal symmetry within the two generations being considered, whereas
the down quark mass matrix has only one eigenvalue nonzero. By selecting this
solution we find that to leading order the third and fourth generation top
quarks and the fourth generation bottom quark are roughly degenerate and the
third generation bottom is massless. As we include the corrections due to the
charged gauge bosons exchange the third generation bottom quark picks up a mass
equal to $I_T$. The corrections to the masses of all other particles are
negligible. Hence we see that there does exist a solution which can accommodate
the experimentally expected value for the third generation quark masses.
However it also predicts an approximately degenerate fourth generation with
mass roughly the same as the third generation top. This possibility is not
ruled out experimentally at present since the current experiments are most
sensitive to
the mass difference within the fourth generation and not to the mass values.
However we should not take this prediction seriously since as discussed
below this prediction will change considerably when we include more than
2 generations.

\subsect{3e.}{Extension to more generations}

The generalization of the above mechanism to more than two generations is
straightforward. Let us consider the case of three generations which may be
taken to be the fourth generation ($T,B$), third generation ($t,b$),  and the
second generation ($c,s$) doublets. The heavy generations will be assumed to
have the
mass pattern that was obtained from consideration of the two generation case.
As discussed above, in order to generate the large mass difference between
the top and the bottom quark it is necessary to select the solution for
which to leading order the fourth generation top $T$, the fourth generation
bottom $B$ and the third generation top $t$ are massive. With this solution
the fourth generation and the third generation top quarks are roughly
degenerate to leading order. Also since the up sector and the down sector have
roughly the same couplings the $T$ and $B$ quarks are also roughly degenerate.
As we will see, the remaining quarks will pick up masses once the charged
boson exchange contributions are included.
To leading order the up sector and down sector quark mass matrices are
\beqn
M_U &=& U_U^\dagger \left(\matrix{m&0&0\cr 0&m&0\cr 0&0&0\cr}\right) U_U
\nonumber \\
M_D &=& U_D^\dagger \left(\matrix{m&0&0\cr 0&0&0\cr 0&0&0\cr}\right) U_D\ \ ,
\eeqn
where $U_U$ and $U_D$ are $3\times3$ unitary matrices and the allowed
generality of different left and right unitary transformations is not necessary
for our discussion. Diagonalization of
these matrices leads to the CKM matrix $U = U_U^\dagger U_D$. Clearly since
$U_U$ and $U_D$ can be arbitrary unitary matrices, the CKM matrix can also be
chosen arbitrarily. We assume that it has the following approximate form,
\beq
U = \left(\matrix{1-\alpha^2/2&\alpha&\alpha\gamma\cr -\alpha& 1-{\alpha^2\over
2}-{\beta^2\over 2}-{\gamma^2\over 2}-\beta\gamma & \beta+\gamma\cr
-\alpha\beta&\beta+\gamma & -1+{\beta^2\over 2}+{\gamma^2\over
2}+\beta\gamma\cr
}\right)\ \ .
\eeq
The matrix $U$ has been obtained by expanding the form given in Ref.
\citenum{chengli} and is unitary up to cubic terms in $\alpha$, $\beta$ and
$\gamma$. We will perform a perturbative calculation of the masses of the light
particles such that it is accurate only to second order in $\alpha$, $\beta$
and $\gamma$ and therefore we do not need any higher order terms in the matrix
$U$.
The corrected mass matrices after including the corrections due to the charged
particle exchanges shown in Fig. 2 can be written as,
\beqn
M_{U1}^{\rm corr} = \left(\matrix{m&0&0\cr 0&m&0\cr 0&0&0\cr}\right) + \delta \
U
\left(\matrix{m&0&0\cr 0&0&0\cr 0&0&0\cr}\right)U^\dagger\nonumber\\
M_{D1}^{\rm corr} = \left(\matrix{m&0&0\cr 0&0&0\cr 0&0&0\cr}\right) + \delta \
U^\dagger
\left(\matrix{m&0&0\cr 0&m&0\cr 0&0&0\cr}\right)U
\eeqn
where $\delta$ is small compared to 1 and can be calculated by explicitly
evaluating the diagram shown in Fig. 2. This result in Eq. 28 is valid at large
enough momenta such that all fermion masses are negligible in comparison to the
momenta flowing in the fermion propagators. To leading order in corrections the
bottom quark picks a mass equal to $m\delta $, all other corrections are higher
order.
The up sector and down sector mass matrices to this order become,
$$\left(\matrix{m&0&0\cr 0&m&0\cr 0&0&0\cr}\right)\ \ ,\left(\matrix{m&0&0\cr
0&m\delta &0\cr 0&0&0\cr}\right)\ \ \mbox{resp.}$$
We now further iterate the fermion mass matrix by reinserting the corrected
mass matrices into diagram shown in Fig. 2. The next iteration yields the
following result for the mass matrices.
\beqn
M_{U2}^{\rm corr} = \left(\matrix{m&0&0\cr 0&m&0\cr 0&0&0\cr}\right) + \delta \
U
\left(\matrix{m&0&0\cr 0&m\delta &0\cr 0&0&0\cr}\right)U^\dagger\nonumber\\
M_{D2}^{\rm corr} = \left(\matrix{m&0&0\cr 0&m\delta &0\cr 0&0&0\cr}\right) +
\delta \ U^\dagger
\left(\matrix{m&0&0\cr 0&m&0\cr 0&0&0\cr}\right)U
\eeqn
It is clear that $\alpha$, $\beta$ and $\gamma$ are still arbitrary. The
eigenvalues of the resulting mass matrices can be easily calculated. The only
nonnegligible corrections are to the $(c,s)$ doublet. Explicit calculation
yields the following values for the smallest eigenvalues,
\beqn
m_{\rm charm}&=&\delta^2(\beta+\gamma)^2m + \delta (\alpha\beta)^2m\nonumber\\
m_{\rm strange}& = &\delta(\beta+\gamma)^2m/2
\eeqn
Therefore we see that the $(c,s)$ has picked up a mass much smaller than the
two remaining generations. However we notice that the strange quark mass
is of lower order in small quantities than the charm quark mass. Therefore
we have the strange quark heavier than the charm, in conflict with
experiments. This problem could have been anticipated. The Feynman diagram
in Fig. 2, which is responsible for the generation of mass of the charm and
strange particles, couples the strange particle to the top and the charm
quark to the bottom. Since the mass of bottom is much smaller than the
top, we end up with a much smaller mass for the charm in comparison to
the strange.

The above problem with the predicted mass values of the $(c,s)$ doublet can
be resolved by including one more generation
 ($T',B'$), making a total of five generations necessary to obtain a realistic
fermion mass spectrum. We again ignore the lightest generation (u,d) for the
present discussion. We now assume that to leading order only the fourth and the
fifth generation down quarks $B$, $B'$ and the fifth generation up quark $T'$
are massive. Then charged gauge particle exchange naturally yield a mass for
the fourth generation top $T$ which is much smaller than the corresponding
bottom
$B$ and give a larger contribution to the third generation top $t$ in
comparison to the corresponding bottom $b$ quark, in direct analogy with the
calculation
including the 3 doublets $(T,B)$, $(t,b)$ and $(c,s)$ described above.
Furthermore now there would be two opposing contributions to the second
generation quarks, charm and strange. The fourth generation will tend to make
charm heavier and the third generation will give larger contribution to the
strange quark. Therefore one has the freedom to adjust the values of the mixing
angles to make the charm quark heavier than the strange quark.

A simple calculation illustrates the basic idea. We now have the fermion mass
spectrum in which the $T'$,
$B'$ and $B$ quarks are nearly degenerate, $T$ is much lighter than
$B$\footnote{Dynamical effects such as $\bar TB$ vector meson bound states must
be included in assesing the impact of a large $T-B$ mass splitting on the
$\rho$-parameter. See for example, P. Q. Hung, Phys. Rev. Lett. {\bf 70}, 888
(1993). A similar dynamical mechanism in technicolor models is presented in R.
Johnson, B.-L. Young and D. McKay, Phys. Rev. {\bf D43} R17 (1991).} and
$t$ and $b$ have even smaller masses with $m_t>>m_b$. The mass matrices can be
written as,
\beqn
M_{U2}^{\rm corr} &=& \left (\matrix{m&0&0&0\cr 0&\delta m&0&0\cr 0&0&\delta_1
m&0\cr 0&0&0&0\cr}\right)\nonumber\\
M_{D2}^{\rm corr} &=& \left (\matrix{m&0&0&0\cr 0&m&0&0\cr 0&0&\delta_2 m&0\cr
0&0&0&0\cr}\right)
\eeqn
where $\delta_2<<\delta_1<<\delta<<1$. This relation can be derived
by keeping only the three heaviest generations and assuming a general form
of the CKM matrix given in Eq. 27.

 We are next interested in
calculating the masses of the $(c,s)$ quark doublet. To simplify the
calculation
we assume that the fifth generation gives negligible contribution to the
strange
and charm quarks. In any case we can make this contribution arbitrarily small
since we have the freedom to adjust the value of $V_{15}$. We therefore include
only the $(T,B)$, $(t,b)$ and $(c,s)$ doublets for our calculation. The third
iteration of the mass matrices then yields,
\beqn
M_{U3}^{\rm corr} = \left(\matrix{\delta m&0&0\cr 0&\delta_1m&0\cr
0&0&0\cr}\right) + \delta\  U
\left(\matrix{m&0&0\cr 0&\delta_2 m&0\cr 0&0&0\cr}\right)U^\dagger\nonumber\\
M_{D3}^{\rm corr} = \left(\matrix{m&0&0\cr 0&\delta_2 m&0\cr 0&0&0\cr}\right) +
\delta \ U^\dagger
\left(\matrix{\delta m&0&0\cr 0&\delta_1 m&0\cr 0&0&0\cr}\right)U\ \ ,
\eeqn
 which leads to the expressions,
\beqn
m_{\rm charm} &=& {\delta\over 2}(\alpha'\beta')^2m \nonumber\\
m_{\rm strange} &=&  \delta^2 (\alpha'\gamma')^2m
+ \delta\delta_1(\beta'+\gamma')^2m
\eeqn
where $\alpha'$, $\beta'$ and $\gamma'$ are the mixings angles for the
CKM matrix at the present order of the calculation. In Eq. 33 only the dominant
terms are shown for the strange and charm mass. We point out that to the
present order we need to have the unitary matrix accurate to fourth order
in the mixing angles and that $\delta_1$ and $\delta_2$ are already of order
($\delta \times$ square of the mixing angles). Therefore Eq. 33 shows that the
charm quark is much heavier than the strange quark.

The process can be continued further to yield the mass of the $(u,d)$ doublet
but the actual calculations get rather cumbersome. At second and higher
iterations it might also be necessary to include other Feynman diagrams besides
the one illustrated in Fig. 2. We plan to present detailed calculations in a
future publication.

The above mechanism can be generalized to leptons and in order for the lighter
charged leptons to acquire masses we need at least one generation with a heavy
neutrino. The W exchange can then generate masses for
the light charged leptons analogously to the case of quarks. In order to be
realistic we again seem to need at least five generations of fermions. To
leading order we assume that only the fourth and the fifth generation neutrino
and the fifth generation electron are massive. Then the mechanism discussed
will lead to a comparatively small mass for the fourth generation electron and
even
smaller masses for the remaining generations. One can see that, in analogy with
the case of the quark mass calculation, the mass of the first three generation
neutrinos will be much smaller than the corresponding charged particles.
However we need to do detailed calculations to determine whether their masses
are
phenomenologically acceptable.

\subsect{3f.} {Goldstone bosons}

A possible objection to the above scenario is that Goldstone bosons
appear if the global horizontal symmetry is dynamically
broken. In order for the model to be realistic we have to generate
masses for these particles. We count the number of Goldstone bosons
present by specializing to the five generation model, which we have argued is
the minimum number of generations required to get a realistic fermion spectrum.
We have broken the global $SU(5)_L\times SU(5)_R$ dynamically. This will lead
to forty eight Goldstone particles. We have also broken the $U(1)_A$ symmetry.
However, in analogy with the resolution to the $U(1)$ problem in QCD, this may
not lead to any Goldstone bosons because of the presence
of the anomaly. Therefore in the case of five generations we have forty eight
massless particles. The generalization of our model to avoid the Goldstone
problem is currently under study.

\sect{4} {Conclusions}

We have presented a simple mechanism for dynamical generation of the fermion
mass hierarchy. The method is implemented within the framework of a left-right
model and is based on the observation that it is possible to generate the CKM
matrix dynamically, if all the fermions are coupled to some
new strong interaction. We find a set of solutions to the gap equation for
which, to leading order only some of the fermion generations becomes massive.
However
due to the generation of mixing angles the lighter generations also pick up
a mass by exchanging charged gauge particles. We have also discussed the
stability of such solutions and have argued that there are no tachyonic bound
states, so these solutions should be
locally stable. For the present paper we have basically emphasized that such
a mechanism is possible without going into a detailed discussion of
phenomenology. We have presented a calculation to indicate that it may be
possible to obtain the experimental values of the fermion mass spectrum if at
least two  heavy generations, namely the fourth and the fifth, are present.
One difficulty that we encountered is the generation of unwanted
Goldstone bosons. This problem has been left unresolved in the present paper.
For this reason it appears that although the model is suggestive
and interesting, in the present form it faces serious phenomenological
problems.
It will clearly be remarkable if the model can be suitably extended to
avoid this problem.

\vskip 1.0cm
\noindent
{\bf Acknowledgements}

We thank Kimball Milton, Herman J. Munczek and John P. Ralston for useful
comments and discussions.
This work has been supported in part by the DOE Grants Nos. DE-FG02-85ER-40214
and DE-FG05-91ER-40636. P.J. thanks Dean Miller for help with the figures.
D.W.M. thanks the Department of Physics at the University of California, Davis,
and in particular Barry Klein, Ling-Lie Chau and Jack Gunion, for hospitality
while this work was being completed.

\vskip 1.0cm
\noindent
\sect{Appendix 1} {Left-Right Symmetry Breaking Lagrangian}

As discussed in section (3a.), the left-right symmetry can be broken by
introducing a new fermion doublet $(\chi_1,\chi_2)$ which couples only to the
right handed and
the $U(1)$ gauge particles. This fermion doublet is assumed to have a
very large mass because of a new interaction which is assumed to be a
vectorial $SU(N)$ interaction. We leave the value of $N$ unspecified in
the present paper. The hypercharge assignment for this doublet in given in
Eq. 21. The lagrangian for this doublet refered to as ${\cal L}(\chi)$ in
section (3a.) can be written as,

\beqn
{\cal L}(\chi) &=& -{1\over 2} tr\ Q_{\mu\nu}Q^{\mu\nu} +
i\left(\matrix{\bar\chi_1&
\bar\chi_2\cr}\right)\slash\partial\left(\matrix{\chi_1\cr \chi_2\cr}\right)
\nonumber\\
&+& g_C\left(\matrix{\bar\chi_1& \bar\chi_2\cr}\right)\slash
C\left(\matrix{\chi_1\cr \chi_2\cr}\right)+g_R\left(\matrix{\bar\chi_1&
\bar\chi_2\cr}\right)_R\tau\cdot\slash A_R\left(\matrix{\chi_1\cr
\chi_2\cr}\right)_R\nonumber\\
&+& \tilde g Y_R\left(\matrix{\bar\chi_1& \bar\chi_2\cr}\right)_R\slash B
\left(\matrix{\chi_1\cr \chi_2\cr}\right)_R + \tilde g\left(\matrix{\bar\chi_1&
\bar\chi_2\cr}\right)_L\left(\matrix{Y_L^U&0\cr 0& Y_L^D}\right)
\slash B\left(\matrix{\chi_1\cr \chi_2\cr}\right)_L
\eeqn
where
$$Q_{\mu\nu} = \partial_\mu C_\nu-\partial_\nu C_\mu - ig_C[C_\mu,C_\nu]\ \ ,$$
$C_\mu$ is the $SU(N)$ gauge field matrix and the fermion field $\chi_i$ also
carry internal index corresponding to this gauge group.
As discussed in section (3a.) if we choose $Y_R$ to be $1/3$ as for all
other quarks, we then have to choose $Y_L^U=4/3$ and $Y_L^D = -2/3$.


\vfill
\eject
\noindent{\bf Figure Captions}
\vspace*{0.8cm}

\noindent {\bf Fig. 1} The Schwinger Dyson equation in the ladder
approximation,
including only the $R_2$ gauge particle exchange, which is expected to give the
dominant contribution. The corrections due to photon, $Z$, $W_L$ and $W_R$ can
be calculated perturbatively.

\vspace*{0.4cm}
\noindent {\bf Fig. 2} The dominant charged gauge particle exchange correction
to the Schwinger-Dyson equation. This correction is responsible for the
generation of mass for the light fermions.

\end{document}